\documentclass[journal]{IEEEtran}
\usepackage{times}
\usepackage{amssymb}
\usepackage{helvet}
\usepackage{mathrsfs}
\usepackage{url}

\usepackage{hyperref}

\usepackage{courier}
\usepackage{booktabs,caption}
\usepackage[flushleft]{threeparttable}
\usepackage{amsmath}
\usepackage{graphicx}
\usepackage{booktabs}
\usepackage{url}	
\usepackage{subfigure}
\usepackage{overpic}
\usepackage{color}
\usepackage[english]{babel}
\usepackage[T1]{fontenc}
\usepackage[utf8]{inputenc}
\usepackage{multirow}
\usepackage{diagbox}
\usepackage{authblk}
\usepackage[numbers, compress]{natbib}
\usepackage{xcolor}

\begin{document}
\title{An Online Platform for Automatic Skull Defect Restoration and Cranial Implant Design}

\author[1,2,*]{Jianning Li}
%\author[3]{Marcell Krall}
\author[1,2]{Antonio Pepe}
\author[1,2,3]{Christina Gsaxner}
%\author[3]{Ulrike Zefferer}
%\author[3]{Gord von Campe}
%\author[3]{Ute Schäfer}
%\author[1]{Dieter Schmalstieg}
\author[1,2,3,*]{Jan Egger}

\affil[1]{Institute of Computer Graphics and Vision, Graz University of Technology.}
\affil[2]{Computer Algorithms for Medicine Laboratory, Graz, Austria.}
%\affil[3]{Department of Neurosurgery, Medical University of Graz, Auenbruggerplatz 29, A-8036 Graz, Austria.}
\affil[3]{Department of Oral and Maxillofacial Surgery, Medical University of Graz.}
\affil[*]{Corresponding authors: jianning.li@icg.tugraz.at, egger@tugraz.at}

\markboth{%IEEE Transactions on Medical Imaging
	}%
{Shell \MakeLowercase{\textit{et al.}}: Bare Demo of IEEEtran.cls for IEEE Journals}

\newcommand{\revise}[1]{{\color{blue}{#1}}}

\maketitle

\begin{abstract}
We introduce a fully automatic system for cranial implant design, a common task in cranioplasty operations. The system is currently integrated in \textit{Studierfenster} (\url{http://studierfenster.tugraz.at/}), an online, cloud-based medical image processing platform for medical imaging applications. Enhanced by deep learning algorithms, the system automatically restores the missing part of a skull (i.e., skull shape completion) and generates the desired implant by subtracting the defective skull from the completed skull. The generated implant can be downloaded in the \textit{STereoLithography} (\textit{.stl}) format directly via the browser interface of the system. The implant model can then be sent to a 3D printer for {\emph{in loco}} implant manufacturing. Furthermore, thanks to the standard format, the user can thereafter load the model into another application for post-processing whenever necessary. Such an automatic cranial implant design system can be integrated into the clinical practice to improve the current routine for surgeries related to skull defect repair (e.g., cranioplasty). Our system, although currently intended for educational and research use only, can be seen as an application of additive manufacturing for fast, patient-specific implant design.
\end{abstract}

\begin{IEEEkeywords}
Cranial implant design, Deep-learning, Cranioplasty, Additive manufacturing, 3D printing, Studierfenster
\end{IEEEkeywords}
\IEEEpeerreviewmaketitle
%---------------------------------------------------------------------------

\section{Introduction}
Cranioplasty refers to the surgical process of repairing skull defects using custom-made cranial implants. Cranioplasty has been known as a costly and time-consuming process due to the bottleneck of the current clinical routine, which primarily relies on high-quality implant design and manufacturing of cranial implants by professional companies. The advancement of fast prototyping technologies such as additive manufacturing (AM) and bio-compatible materials has facilitated fast and low-cost manufacturing of medical implantable devices. However, a solution for rapid and low-cost implant design is still missing. In the clinics, the cranial implants are currently designed by professional designers -- e.g. contractors -- with the aid of commercial solutions, which represents a costly and time-consuming operation. For example, in a case study by \cite{casestudy},
the cranial implant for a patient going through 
brain tumor surgery was designed by a professional design research center in the UK, whereas the patient was from Spain. The patient's computed tomography (CT) scan was transferred from the Spain hospital to the design center in the UK. Several commercial software, such as the  MIMICS (Materialise NV, Belgium) and Geomagic (3D Systems, South Carolina) were involved in the cranial implant design process. After the design is finished, the implant was manufactured by another company based in the UK, using metal 3D printing (titanium). After the cranial implant was manufactured, it is sent back to the hospital in Spain so that the responsible neurosurgeon can perform the cranioplasty on the patient. 

Therefore, the optimization of the current workflow in cranioplasty remains an open problem, with implant design as primary bottleneck.

In this study, we introduce a fast and fully automatic system for  cranial implant design. The system is integrated in a freely accessible online platform. Furthermore, we discuss how such a system, combined with AM, can be incorporated into the cranioplasty practice to substantially optimize the current clinical routine.

Initiated by two students from Graz University of Technology \cite{Max2019, Daniel2019}, Studierfenster (\url{www.studierfenster.at}) is a cloud-based, open-science platform for medical image processing, which can be accessed via the browser. Multiple additional features have been integrated into the platform since its first release, such as 3D face reconstruction from a 2D image, inpainting and restoration of aortic dissections (ADs) \cite{Prutsch2020DesignAD}, automatic aortic landmark detection and automatic cranial implant design. Most of the algorithms behind these interactive features run on the server side and can be easily accessed by the client using a common browser interface. The server-side computations allow the use of the remote platform also on smaller devices with lower computational capabilities.  

Our automatic cranial implant design system has been incorporated into Studierfenster since the early development phase and has been thereafter iteratively optimized. Even if the details of the system are not covered in this study, we point out that the problem can be formulated as a volumetric shape completion task. A correct problem formulation is essential for solving this challenging task of automatic cranial implant design.   

\begin{figure*}[ht]
     \centering
        \includegraphics[width=\textwidth]{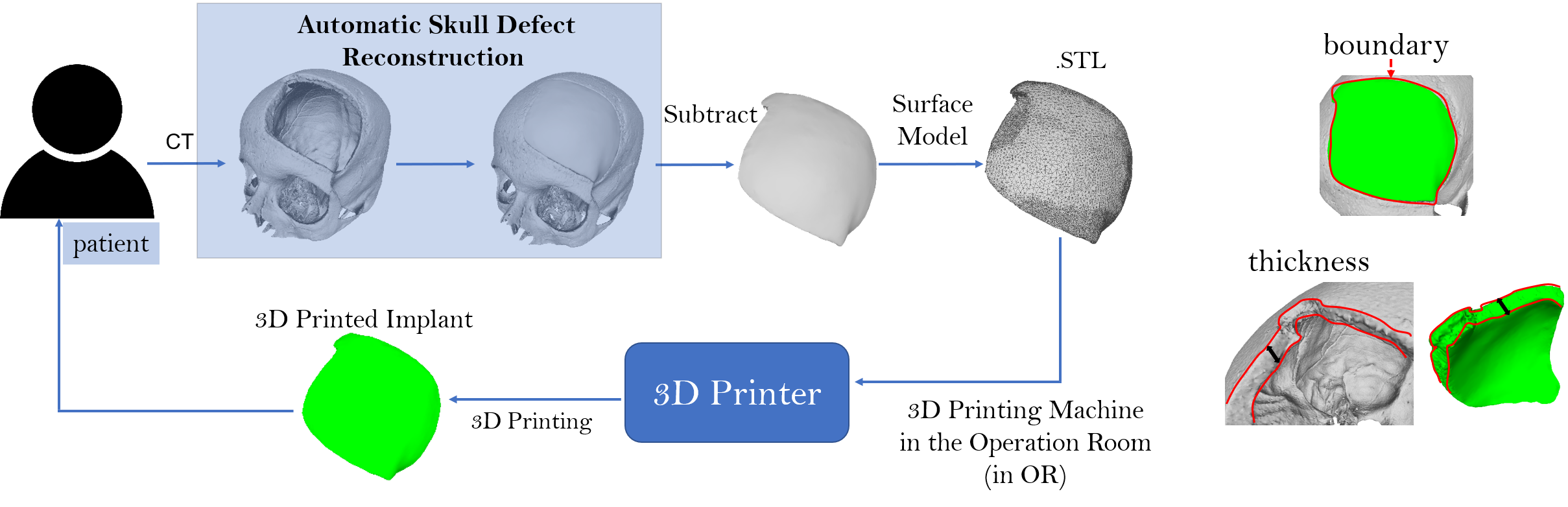}
     \caption{Illustration of the \emph{In-Operation Room} (in-OR) process for cranial implant design and manufacturing. Left: a possible workflow. Right: how the implant should fit with the skull defect in terms of defect boundary and bone thickness.}
     \label{fig:fig2}
\end{figure*}

%---------------------------------------------------------------------------
\section{Cranial Implant Design: Current Clinical Routine}
Cranial implant design is usually associated with skull defect repair, a surgical procedure usually known as cranioplasty. Various events can lead to a skull defect, such as head trauma or previous surgeries related to skull deformity correction or brain tumor removal, during which the surgeons need to remove a part of the cranial bone to access the brain area. To repair the defect after a surgery, a synthetic substitute for the removed bony structure is needed. This synthetic substitute is usually created with of titanium or bio-compatible polymers as the removed bony structure is subject to physical damage and contamination and it is therefore not reusable.       

In the surgical scenario of brain tumor removal, after the completion of the operation, the postoperative head CT scan of the patient is sent to a third-party manufacturer. The manufacturer segments the incomplete skull in the head CT scan and, based on the segmentation, designs a patient-specific cranial implant \cite{casestudy}. In particular, this process can be summarized as follows:
\begin{itemize}
    \item imaging data acquisition (e.g, head CT),
    \item skull segmentation in the imaging data,
    \item conversion of the skull to a 3D CAD model,
    \item patient-specific implant design based on the skull model.
\end{itemize}

Due to the high requirements for cranial implant design, such as the professional experience required and the commercial software, cranioplasty can result in a costly operation for the health care system. On top, the current process is a cause of additional suffering for the patient, since a minimum of two surgical operations are involved: the craniotomy, during which the bony structure is removed, and the cranioplasty, during which the defect is restored using the designed implant. When the cranial implant is externally designed by a third-party manufacturer, this process can take several days \cite{casestudy}, leaving the patient with an incomplete skull.

%\begin{figure*}[ht]
%     \centering
%        \includegraphics[width=\textwidth]{figures/fig3.png}
%     \caption{}
%     \label{fig:fig3}
%\end{figure*}

%---------------------------------------------------------------------------
\section{Low-cost, Fast and \emph{In-Operation Room} (in-OR) Cranial Implant Manufacturing}
As previously introduced, the known limitations of the current routine for cranial implant design include costly, time-consuming and out-of-the-operation-room operations. Researchers have been looking for solutions to overcome these shortcomings. One of the suggested solutions is to develop \emph{ad hoc} free CAD software for cranial implant design \cite{Gallinproceedings, Chenarticle, Janarticle, Marzolaarticle, Marreiros2016CustomID}. However, even if the introduction of \emph{ad hoc} free CAD software can potentially reduce the related costs, the design process is still time-consuming and requires expertise. Therefore, a low cost, fast (e.g., fully automatic) and on-site design and manufacturing of cranial implants remains an open problem and a promising direction worthy of more attention.

AM offers the opportunity of fast manufacturing of 3D models and has been successful in various medical applications \cite{Salmi2013AccuracyOM,Javaid2018AdditiveMA,Dawood20153DPI,Javaid2019CurrentSA,Haleem2018AdditiveMA,Youssef2017AdditiveMO}, including the manufacturing of 3D cranial implants \cite{Modi2018DesignAA, Jardini2014CranialR3, Park2016CranioplastyEB,Kim2012CustomizedCI,MoralesGmez2018CranioplastyWA,Bonda2015TheRR}. Using AM facilitates fast, low-cost and \emph{in loco} manufacturing of cranial implants but needs to be combined with a fast and fully automatic solution for implant design.

\autoref{fig:fig2} shows the optimized workflow for cranial implant design and manufacturing, combining AM (3D printing) with a fully automatic solution for implant design. After a portion of the skull is removed by a surgeon, the skull defect is reconstructed by a software given as input the post-operative head CT of the patient. The software generates the implant by taking the difference between the two skulls. Afterwards, the surface model of the implant is extracted and sent to the 3D printer in the operation room for 3D printing. The implant can therefore be manufactured \emph{in loco}. The whole process of implant design and manufacturing is done fully automatically and in the operation room.  

In comparison with the traditional cranial implant design and manufacturing workflow, we summarize the optimized procedure as follows:

\begin{itemize}
    \item Imaging data acquisition (e.g, head CT),
    \item Skull extraction from imaging data,
    \item Fully automatic implant modeling,
    \item On-site 3D printing (manufacturing) of the implant model. 
\end{itemize}

\begin{figure*}[h]
     \centering
        \includegraphics[width=\textwidth]{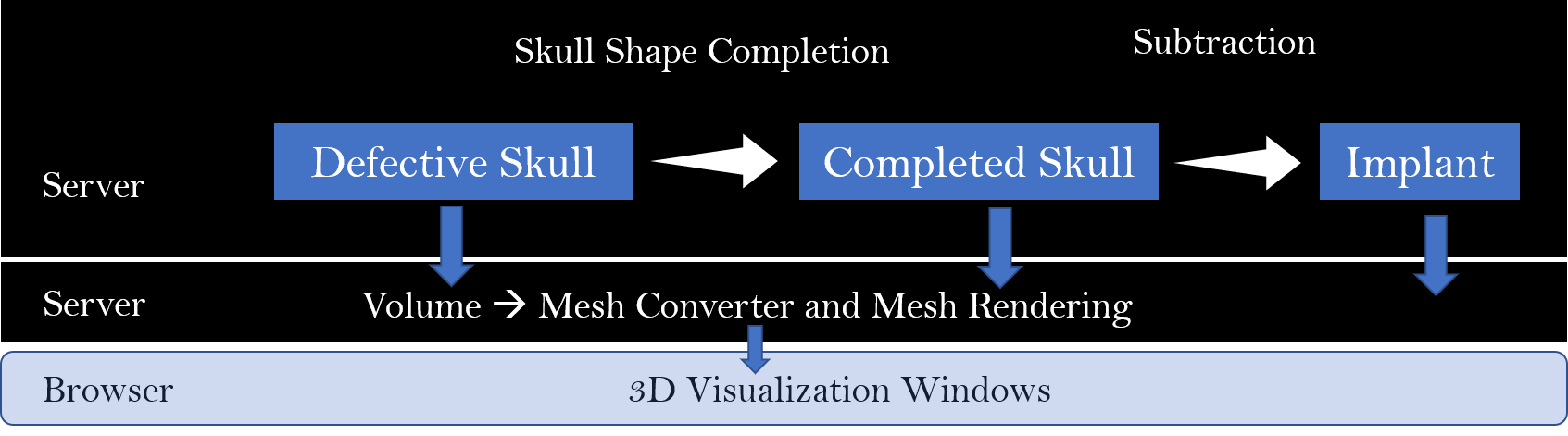}
     \caption{The architecture of the automatic cranial implant design system in Studierfenster. The server side is responsible for implant generation and mesh rendering. The browser side is responsible for 3D model visualization and user interaction.}
     \label{fig:fig1}
\end{figure*}

The optimized procedure can significantly improve the entire surgical process. First, as no expert and commercial software is needed for the implant design, the cost can be significantly reduced. Second, enhanced by the fully automatic implant design software and AM, the waiting time for the implant can be decreased substantially, therefore reducing the suffering of the patient. Cranioplasty can be performed shortly after tumor removal. Third, the implant can be designed and manufactured in the operation room, without the need for external suppliers.      

Based on this considerations, we can see that the automated system for cranial implant design is the key component to an optimized surgical procedure. The automated design remains, however, a very challenging task, as it involves many considerations. First, as the implant should provide a protection to the brain, its shape has to fit precisely within the defected region on the skull, which includes the boundary of the defect and the thickness of the skull surface (\autoref{fig:fig2}, right). Second, the shape of the implant should be consistent with the skull. Even if the geometric shape of human skulls is generally not complex, the irregular defect on the skull can come in various dimensions, shapes and positions. It is, therefore, still challenging to generate the implant in an automated manner, which can satisfy the criteria for boundary, thickness and shape consistency.  
 
%---------------------------------------------------------------------------
\begin{figure*}[h]
     \centering
        \includegraphics[width=0.75\textwidth]{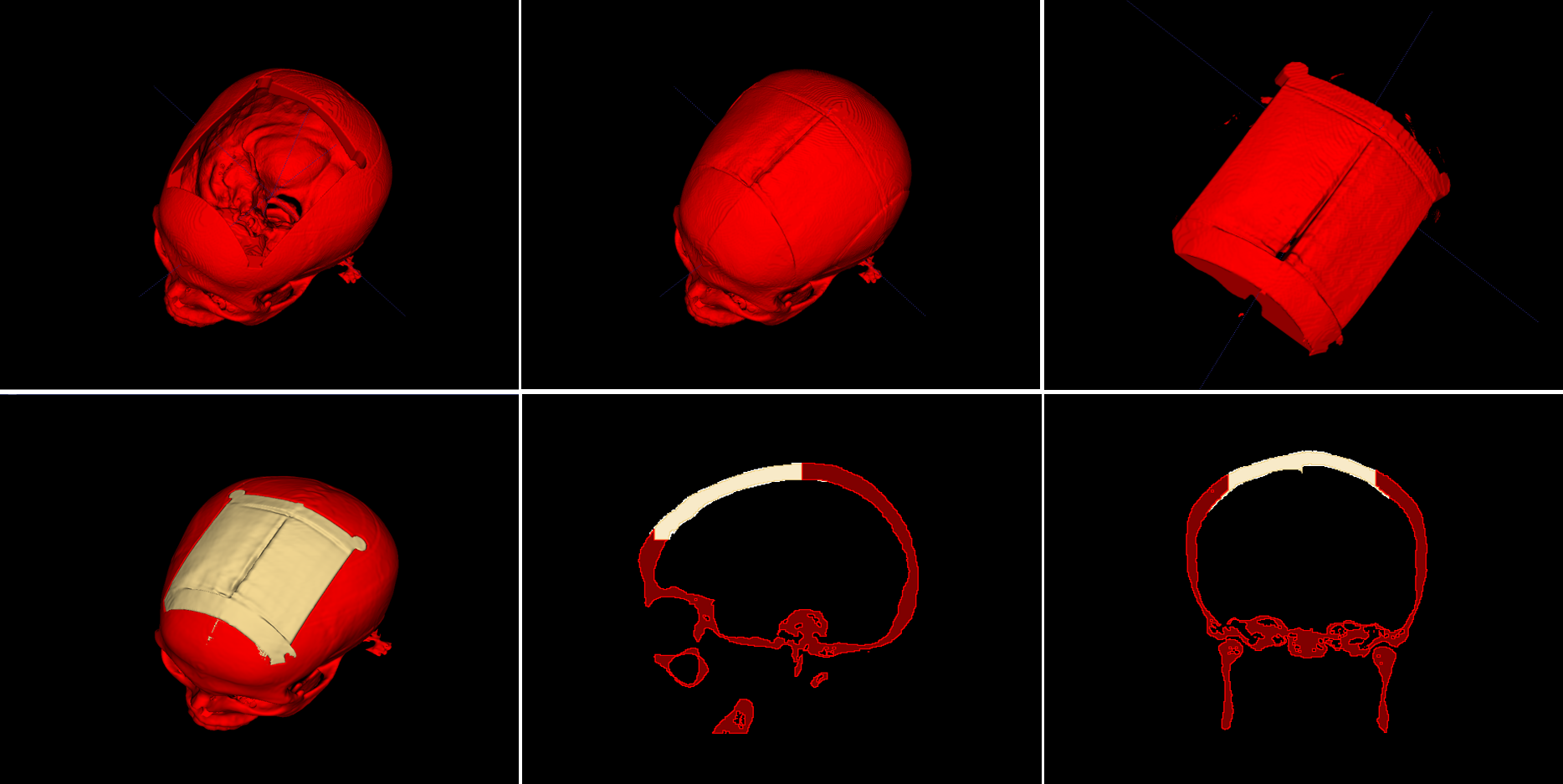}
     \caption{An example of automatic skull defect restoration and implant design. First row: the defective skull, the completed skull and the implant. Second row: how the implant fits with the defective skull in term of defect boundary, bone thickness and shape. To differentiate, the implant uses a different color from the skull.}
     \label{fig:fig3}
\end{figure*}

\section{the Automatic Cranial Implant Design Module in Studierfenster}
Focusing on an automated and freely accessible solution for cranial implant design, we have developed a deep learning-based algorithm for skull defect restoration and cranial implant model generation. The algorithm first completes a defective skull and then generates the implant by taking the difference between the completed skull and the defective skull. The idea is similar to that of Morais et al. \cite{Morais2019}, but with substantial improvements. For instance, in the previous work, the skull dimension was restricted to $30^3$, $60^3$ and $120^3$. However, the imaging data acquired in clinical routine is usually of high dimension such as $512 \times 512 \times Z$, $Z$ being the number of axial slices. Our algorithm can process the high dimensional imaging data directly.   

To ease the accessibility, we have integrated the algorithm into Studierfenster, so that users can interact with the algorithm via a browser using a standard internet connection. 
\autoref{fig:fig1} shows the architecture of the implant design system in Studierfenster. On the server side, the algorithm receives as input a defective skull volume and then the skull completion process is started. Finally, the algorithm subtracts the defective skull volume from the reconstructed skull volume. Note that the algorithm processes volumes instead of a 3D mesh model. For visualization purposes, an additional algorithm, also on the sever side, converts these volumes (the defective, completed skull and the implant) into 3D surface mesh models using the STereoLithography (.stl) format. After rendering, these models, which are also downloadable in the \emph{.stl} format, are shown in the browser window for inspection and verification by the user.  

The usage of the system is summarized as follows:
\begin{itemize}
    \item Access Studierfenster (\url{http://studierfenster.tugraz.at/}) and press the \textit{Implant Generation} button under \textit{3D Skull Reconstruction}.   
    \item In the first window, press \textit{Choose File} to select the defective skull (in the \emph{.nrrd} format) and then press \textit{Upload and Reconstruct}: the data will be uploaded and the 3D defective model will be rendered in the first window.
    \item In the second window, press \textit{Choose File} to select the same defective skull and then press \textit{Upload and Reconstruct}: the defective skull will be completed and shown in the second window.
    \item After the first two steps are finished, press \textit {Start Generation} in the third window: the implant will be generated and shown in the corresponding window. 
    \item Download the 3D implant model (in .stl format) for post-processing if necessary (optional).
\end{itemize}

A YouTube video shows a demonstration of the system usage: \url{http://y2u.be/pt-jw8nXzgs}.

As introduced, the system is easily accessible and allows a fast and fully automatic design of cranial implants. The system can be easily integrated into the optimized surgical procedure discussed in Section \textbf{III}. 

In \autoref{fig:fig3}, we show an example of automatic skull defect restoration and cranial implant design. The head CT, which is in DICOM format, is selected from the public dataset QC500 (\url{http://headctstudy.qure.ai/dataset}). The skull is segmented from the head CT using thresholding (from 150 HU to maximum) and a large, artificial defect is injected into the skull to create a defective skull.  As can be seen in \autoref{fig:fig3}, the algorithm is able to complete the defective skull automatically. The implant is obtained by subtracting the defective skull from the completed skull.
\autoref{fig:fig3} (the second row) also shows that the implant can fit precisely with the defective skull in term of defect boundary, bone thickness and shape consistency.

\section{Description of the Implant Generation Algorithm}
The automatic skull defect reconstruction in \autoref{fig:fig2} is being formulated as a volumetric shape completion problem, where a defective skull shape $\mathbf{S}_{d}$ is completed automatically. The implant $\mathbf{I}$ can be obtained by taking the difference between the completed skull shape $\mathbf{S}_{c}$ and the defective skull shape:

\begin{equation}
\mathbf{I}=\mathbf{S}_{c}-\mathbf{S}_{d} 
\label{eq:implant}
\end{equation}

Equation ~\eqref{eq:implant} is the key element of our problem formulation, even though various approaches can be used to reconstruct $\mathbf{S}_{c}$ from $\mathbf{S}_{d}$. 
Alternatively, the problem can also be formulated to reconstruct $\mathbf{I}$ directly from $\mathbf{S}_{d}$:

\begin{equation}
\mathbf{I}=\mathbf{R} \cdot \mathbf{S}_{d} 
\label{eq:direct}
\end{equation}

where $\mathbf{R}$ is the reconstruction matrix. This formulation avoids the intermediate step to reconstruct the entire skull. Instead, it reconstructs the implant directly. Similarly, there are various approaches to construct the reconstruction matrix $\mathbf{R}$.

In our earlier studies, we have demonstrated that both types of formulation are effective in solving the problem.

\section{Conclusion and Future Work}
In this paper, we introduced our online system for automatic cranial implant design, which, combined with additive manufacturing, can substantially optimize the current clinical routine for cranioplasty. The system is currently intended for educational and research use only, but represents the trend of technological development in this field. As the system is integrated in the open platform Studierfenster, its performance is significantly dependent on the hardware/architecture of the platform. The conversion of the skull volume to a mesh can be slow, as the mesh is usually very dense (e.g., millions of points). This will be improved by introducing better hardware on the server side. Another limiting factor is the client/server based architecture of the platform. The large mesh has to be transferred from server side to browser side in order to be visualized, which can be slow, depending on the quality of the user's internet connection.

%\section*{Competing interests}
\section*{Acknowledgment}
This work receives the support of CAMed - Clinical additive manufacturing for medical applications (COMET K-Project 871132, see also \url{https://www.medunigraz.at/camed/}), which is funded by the Austrian Federal Ministry of Transport, Innovation and Technology (BMVIT), and the Austrian Federal Ministry for Digital and Economic Affairs (BMDW), and the Styrian Business Promotion Agency (SFG). Further, this work received funding from the Austrian Science Fund (FWF) KLI 678-B31 (enFaced - Virtual and Augmented Reality Training and Navigation Module for 3D-Printed Facial Defect Reconstructions) and the TU Graz Lead Project (Mechanics, Modeling and Simulation of Aortic Dissection).

%---------------------------------------------------------------------------

%\newpage
%\clearpage
\bibliographystyle{IEEEtranN}
\bibliography{refs}
\ifCLASSOPTIONcaptionsoff
  \newpage
\fi

\end{document}